\documentclass[
 amsmath,amssymb,
 aip,
reprint,
]{revtex4-1}

\usepackage{graphicx}
\usepackage{dcolumn}
\usepackage{bm}
\usepackage[utf8]{inputenc}
\usepackage[T1]{fontenc}
\usepackage{mathptmx}

\usepackage{CJKutf8}
\usepackage{wasysym}
\usepackage{bbold}
\usepackage{xcolor}
\usepackage[colorlinks=true,allcolors=blue]{hyperref}%
\begin{document}

\title[Mid-infrared photoconductivity spectra of single tellurium nanowires]{Mid-infrared photoconductivity spectra of single tellurium nanowires}

\author{Pengke Li 
}
\email{pengke@umd.edu}
\affiliation{Department of Physics,
U. of Maryland, College Park, MD 20742}
\date{\today}
\begin{abstract}
Due to its low bandgap and high optical efficiency, tellurium is considered an important material candidate for mid-infrared applications. Taking advantage of its structural anisotropy, we fabricated tellurium nanowire devices and investigated the radiative interaction of charge carriers by the polarization-resolved photoconductivity spectra under mid-infrared illumination. The intensity of the photoresponse shows sensitive dependence on temperature and could be significantly boosted by positive voltage bias from the back gate.
\end{abstract}

\maketitle  
\section{Introduction}

\begin{figure}
\includegraphics[width=3.in]{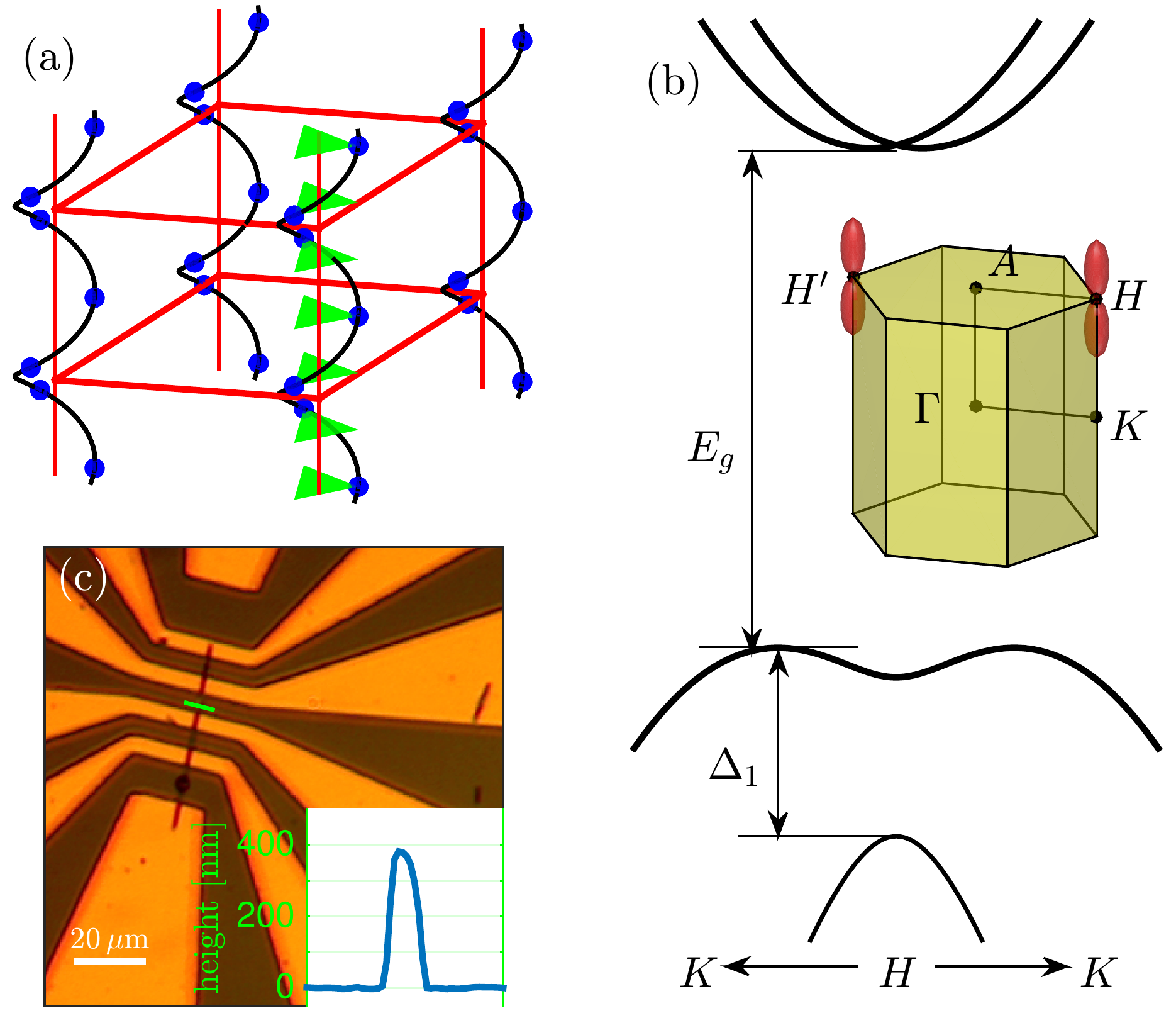}
\caption{ (a) Te lattice composed of 3-fold helical chains in a trigonal structure.  The equilateral triangles (green color) highlight Te atoms (blue dots) on a single helical chain (in black). The parallelepiped with red edges represents a unit cell. (b) Te band structure close to the gap region along the $K-H-K$ direction (see the inset of the Brillouin zone, with the valence band valleys highlighted in red). The gap-edge conduction and valence bands are emphasized by bold lines. The quasi-direct bandgap is $E_g=0.335$~eV, and the split-off gap due to the spin-orbit coupling between the two valence bands is $\Delta_1 = 0.11$~eV. (c)  Optical image of a 50~$\mu$m long TeNW device, with thickness $\sim$380~nm as shown by the AFM profile (for the path in green) in the inset.
\label{fig:f1} }
\end{figure}

Single-crystal elemental tellurium (Te) is composed of three-fold helical atomic chains organized in a trigonal configuration [see Fig.~\ref{fig:f1}(a)]. 
The absence of inversion symmetry in this lattice (as that of selenium) is distinct from most other elemental materials.\cite{Loferski_PR1954} Atoms on each helical chain are valence-bonded, whereas neighboring chains are weakly coupled by van~der~Waals interaction. This unusual anisotropic nature of the uniaxially strong bonding along the $c$-axis of the trigonal lattice motivates recent efforts of mechanical exfoliation of one dimensional (1D) Te atomic chains\cite{Churchill_NRL2017} [similar to the isolation of two dimensional (2D) graphene from bulk graphite], as well as theoretical explorations on the edge states of helical atomic chains\cite{Pengke_PRB2017} and their dispersion into 2D bands on the surface normal to the $c$-axis.\cite{Pengke_PRB2018}

The anisotropy of the Te lattice structure also gives rise to various anisotropic optical properties as governed by the unique electronic structure. One of the most well-known consequences is the polarization dependence of excitation across the fundamental bandgap of Te, as reported in early theoretical\cite{Callen_JCP1954,Dresselhaus_PR1957} and  experimental\cite{Loferski_PR1954,Tutihasi_PR1969} studies on bulk optical absorption. Te is a semiconductor with a room temperature quasi-direct bandgap $E_g=$~0.335~eV. Due to the strong spin-orbit coupling (SOC) of the heavy nuclei (Te Z=52), the gap-edge crystal momenta of the conduction and valence bands are very close but shifted away from the time-reversal pair of $k$-points $H$ and $H'$ at the corners of the Brillouin zone [see Fig.~\ref{fig:f1}(b)]. \cite{DOI_JPSJ1970} Specifically, SOC between the conduction band states is $k$-dependent and vanishes at $H$ (as characterized by the double degeneracy), whereas the valence bands is dominated by $k$-independent SOC that breaks the degeneracy at the $H$-point, resulting in a shallow camelback-shape valence band edge. Right at $H$, symmetry excludes the cross-bandgap optical transition with polarization of the light longitudinal to the $c$-axis of the crystal, and only the transverse polarized light is symmetry-allowed. Away from the $H$ point, however, the mixing of components from remote bands due to $k\cdot p$ perturbation opens higher-order channels for optical transition with longitudinal polarization.
In addition to the strong influence on the  energy dispersion, SOC is also responsible for the special chiral properties of the spin-dependent eigenstates, determined by the chirality of the helical chains of the lattice. 
For example, the correlation between the spins and the crystal momenta along the $c$-axis governs the optical orientation selection rules, leading to the the circular photogalvanic effect\cite{Asnin_SSC1978}
when interacting with circularly polarized light. Recent theoretical study on the first principle bandstructure of Te predicted Wely nodes close to the bandgap carrying topologically nontrivial spin-texture that could evolve under appropriate pressure condition.\cite{Hirayama_PRL2015}
 
The optical efficiency of Te is high, \cite{Edwards_SSE1961} thanks to the quasi-direct bandgap (corresponding to a wavelength $\sim$ 3.7 $\mu$m),
making the material an attractive candidate for various mid-infrared optical applications, such as photon detection,\cite{Edwards_IP1961} gas molecule sensing,\cite{Tsiulyanu_book2011} mid-infrared lasing,\cite{Choi_JPCL2019} etc. Previous studies in this spectral range focused on p-type bulk samples, in which Te vacancies introduce a hole density level of $\sim 10^{14}$~cm$^{-3}$ even for the purest samples.\cite{Tutihasi_PR1969} Due to these residual carriers in the bulk, photocurrent generated from absorption (mostly at the surface) is usually accompanied by relatively large dark current in the bulk,\cite{Loferski_PR1954} limiting the possible applications in practice. 
Further experiments demonstrated enhanced relative photoconductivity (``detectivity'') with decreasing sample thickness down to several hundreds of microns.\cite{Shyamprasad_IP1981,Shih_IP1981} 
Presumably, if the material is in a dimension smaller than the absorption length, the entire body of the sample could be optically excited, which maximize the absorption efficiency. 
Given the lattice anisotropy in favor of stronger chemical bond along the atomic chain ($c$-axis), the nanowire format of Te is naturally preferable to be synthesized to meet this requirement.
A nanowire device on a dielectric substrate also provides a control via back gate voltage, which could (i) repel intrinsic carriers and reduce the dark current, and (ii) spatially separate the photo-excited electron-hole pairs so that the carrier recombination lifetime is prolonged and the photoconductivity is enhanced. 
Moreover, a Te nanowire (TeNW) with thickness comparable to the optical wavelength could behave as an optical cavity confining photons in resonance modes,\cite{Cao_NM2009} which further promotes optical excitation in the material within certain wavelength ranges.

With these motivations, we present in this paper a comprehensive study of the photoconductivity of single-TeNW devices in the mid-infrared range between 2.4~$\mu$m to 4~$\mu$m, focusing on direct band-to-band transition of the gap edge conduction and valence bands.
We fabricated photoconductive devices based on chemically synthesized TeNW samples, as shown by the example in Fig.~\ref{fig:f1}(c). 
The polarization-resolved photoconductivity spectra is found to be sensitive to the temperature condition and could be strongly enhanced by the gate bias. 
The observed spectra of the devices show inconsistent features compared with our theoretically calculated resonant modes based on the Mie formalism. The possible origins of such a discrepancy is analyzed.

\section{Experimental details}
Hydrothermal method was used for the synthesis of TeNWs.\cite{Luo_JNR2012} In a typical synthesis, chemicals including 70~mg Na$_2$TeO$_3$ and 80~mg Na$_2$S$_2$O$_3$ are dissolved in 13~mL DI water. The solution is put in a 25~mL Teflon container and sealed with a stainless steel autoclave. After being heated at 180$^\circ$C for 20 hours, the autoclave is cooled in air. The product solution contains suspended solid TeNWs reduced from Na$_2$TeO$_3$. The solution is left still to let the solid TeNWs precipitate at the bottom, and then the liquid is replaced by DI water. This washing process is repeated several times. X-ray diffraction measurement of the nanowires shows peaks corresponding to the trigonal lattice structure of Te crystals.

To fabricate single nanowires electronic devices, TeNWs are dispersed on p-type Si wafer chips with 100~nm thermally-oxidized layer. Conventional photolithography is used to make contact patterns on the chosen nanowires. Metallization is done by electron beam evaporation of Ti (1 nm) and Au ($\sim$70 nm). During the deposition of Au, the substrate is tilted to ensure continuous metal coverage on the sidewall of the nanowire. Robust yield of ohmic contacts was achieved using this procedure, with relatively small contact resistance.  Fig.~\ref{fig:f1}(c) shows the optical image of a typical device on a $\sim$50 $\mu$m long TeNW with 380 nm diameter given by the atomic force microscope (AFM). 

The devices are mounted in a cryostat cooled by flowing nitrogen gas. DC resistance is measured by a source-measure unit (Keithley 2400), with back gate bias voltage applied between one of the metal contacts and the silicon substrate. The resistivity of the TeNW material at zero gate bias in the dark is around $\sim 0.1\,\Omega\cdot$cm at room temperature, and decreases monotonically when cooled down to 77~K, by a factor of around 2 to 4 varying among different samples.
Back gate bias measurement confirms that the TeNW material is p-type, consistent with bulk-grown samples in which residual hole carriers are introduced by Te vacancies. Note that among different samples, the gate-tunability of the DC resistance varies with the thickness of the nanowire: thinner samples are affected more by gate bias than thicker ones.

Our optical setup was designed to measure the photoconductivity spectrum of the TeNW devices. The light source is a broadband infrared globar (Thorlabs SLS203L) whose emission is longpass filtered with cut-off wavelength at 2.3 $\mu$m. After an optical chopper, the light is dispersed by a monochromator (Jarrell-Ash 82-997) with a 150 lines/mm grating. The monochromator output light is polarized by a Brewster polarizer made of four parallel intrinsic Si wafers providing an extinction ratio $>10000:1$, and is then focused on the sample stage in the cryostat with $c$-cut sapphire windows (to eliminate spurious birefringence). The spectrum of the optical system is calibrated by an InAsSb photovoltaic detector (Hamamatsu P13243). The power density of the incident light is around 0.3~mW/cm$^2$ at the $\approx 3$mm$^2$ focus on the cryostat.  

\begin{figure}
\includegraphics[width=3.25in]{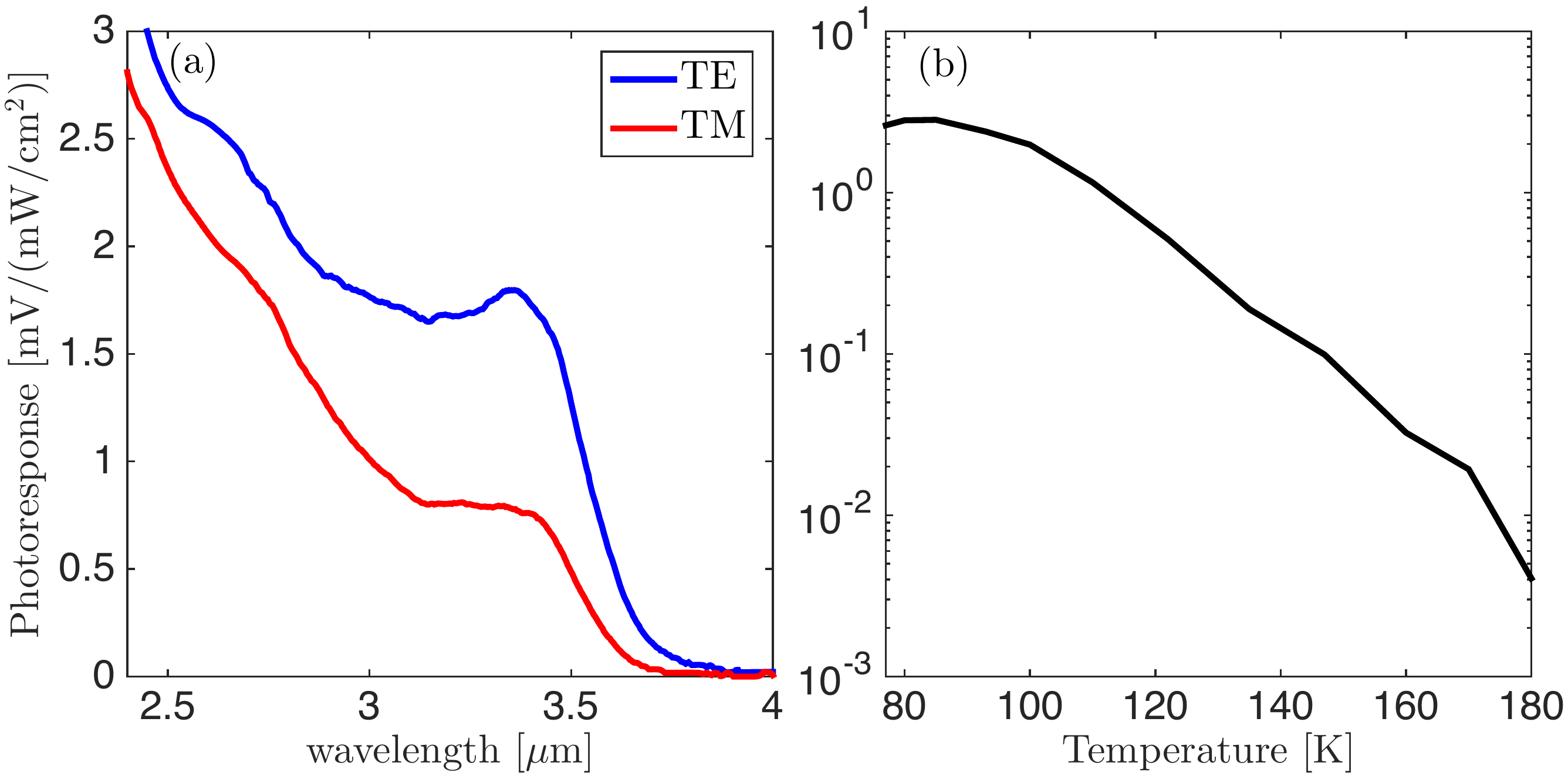}
\caption{\label{fig:temp_dep} 
(a) Mid-infrared photoresponse spectra of a 900~nm-thick TeNW device, under transverse electric (TE, electric field perpendicular to the $c$-axis) and transverse magnetic (TM, electric field parallel to the $c$-axis) polarized illuminations. (b) Amplitude of the TE photoresponse at 2.8~$\mu$m verses temperature.}
\end{figure}

With the TeNW devices biased by constant electric current (on the order of 10~$\mu$A), the photoresponse, i.e. the light-induced variation of the resistance is measured by the AC voltage across the sample, using a lock-in amplifier (SRS SR830) at a chopper frequency of 23 Hz. Throughout this study, the maximum lock-in readout voltage is $\sim1\%$ of the DC voltage, indicating 1. small-injection condition of the light-induced excess carriers,\cite{Vis_JAP1964} and 2. the photoconductivity is inversely proportional to the measured photoresistivity.

\section{Mid infrared photoresponse and temperature dependence}

In Fig.~\ref{fig:temp_dep}(a), we present the polarization-resolved photoresponse spectra of a TeNW device at 77K. Here, by convention, `TE' and `TM' represent linear polarizations of the incident light being transverse and longitudinal to the nanowire axis, respectively. It is clearly seen that the TE spectrum demonstrates stronger amplitude close to the cut-off wavelength around $\sim 3.7$~$\mu$m, compared with the TM spectrum. This is consistent with previous bulk absorption experiments\cite{Loferski_PR1954,Tutihasi_PR1969} and theoretical analyses\cite{Callen_JCP1954,Dresselhaus_PR1957} showing that the optical transition at the forbidden band edge is driven by the TE light at the lowest order, whereas TM light is responsible for higher order and weaker transition close to the bandgap. 

The relative shape of the photoresponse spectra do not change with elevated temperature, whereas their amplitudes is significantly suppressed, which provides information about the recombination mechanisms of excess carriers in the material. Fig.~\ref{fig:temp_dep}(b) presents the amplitude of the TE spectrum at 2.8~$\mu$m in the temperature range between 77~K and 180~K. The photoresponse varies relatively slowly at low temperature, but then undergoes a sharp decrease beyond $\sim100$~K. Such a temperature dependence is shared by all the TeNW samples.

The photoconductivity $\Delta\sigma$ depends on the density of excess carriers and their mobility as
$\Delta\sigma = e(\Delta n \mu_n+\Delta p \mu_p)$,
in which $e$ is the elementary charge. Here, we have $\Delta n = \Delta p \sim G\tau$ for the density of electron and hole pairs generated by illumination, with $G$ the generation rate and $\tau$ the total recombination lifetime. 
Since the carrier mobilities $\mu_{n,p}$ of our devices have relatively weak temperature dependence (as evidenced by the small variation of the dark resistivity), the strong suppression of the photoconductivity at elevated temperature is mostly attributed to the enhanced recombination rate $\tau^{-1} = \tau_\text{rad}^{-1}+\tau_\text{SRH}^{-1}+\tau_\text{Auger}^{-1}$, where we consider all three major recombination mechanisms in semiconductors: radiative, trap-assist, and Auger. 

Band-to-band radiative recombination is strong in regular quasi-direct low band gap semiconductors. In our p-type TeNW samples, the radiative recombination lifetime $\tau_\text{rad} \approx (Bp_0)^{-1}$, in which $p_0$ is the hole density at equilibrium and is weakly dependent on temperature in our range of interest. $B\propto(k_BT)^{-3/2}\exp(E_g/k_BT)$ is the radiative capture probability\cite{Li_SPE} that becomes stronger at lower temperature. As a result, the reduction of radiative recombination lifetime is responsible for the saturation of the photoresonse of the TeNW devices below 100~K.

Secondly, our nanowire samples are unintentionally p-type due to the synthesis process, with fairly high concentration of defects and vacancies responsible for nonradiative trap-assisted recombination. According to the Shockley-Read-Hall theory, in the case of hole-doped semiconductor, the trap-assisted recombination lifetime $\tau_\text{SRH}$ is governed by the minority carrier (electron) lifetime $\tau_n$ which decreases as temperature increases following $\tau_n\propto T^{-3/2}$ at low $T$,\cite{Schenk_SSE1992} due to higher probability of thermal capture by the recombination center.

Thirdly, the low bandgap and large carrier density of our samples suggest a large role for Auger recombination. For p-type tellurium, the Auger lifetime is $\tau_\text{Auger} \approx(p_0^2C_p)^{-1}$. Here, $C_p$ is the capture coefficient due to hole-hole collision with a temperature dependence $C_p\propto T^{3/2}\exp(-E_g/2k_BT)$.\cite{Li_SPE} As a result, $\tau_\text{Auger}$ decreases with increasing temperature, similar to $\tau_\text{SRH}$, but opposite to $\tau_\text{rad}$. Thus both the  trap-assisted and Auger recombinations could be responsible for the significant suppression of the photoresponse beyond 100~K.

\section{Gate bias dependence}

\begin{figure}
\includegraphics[width=3.3in]{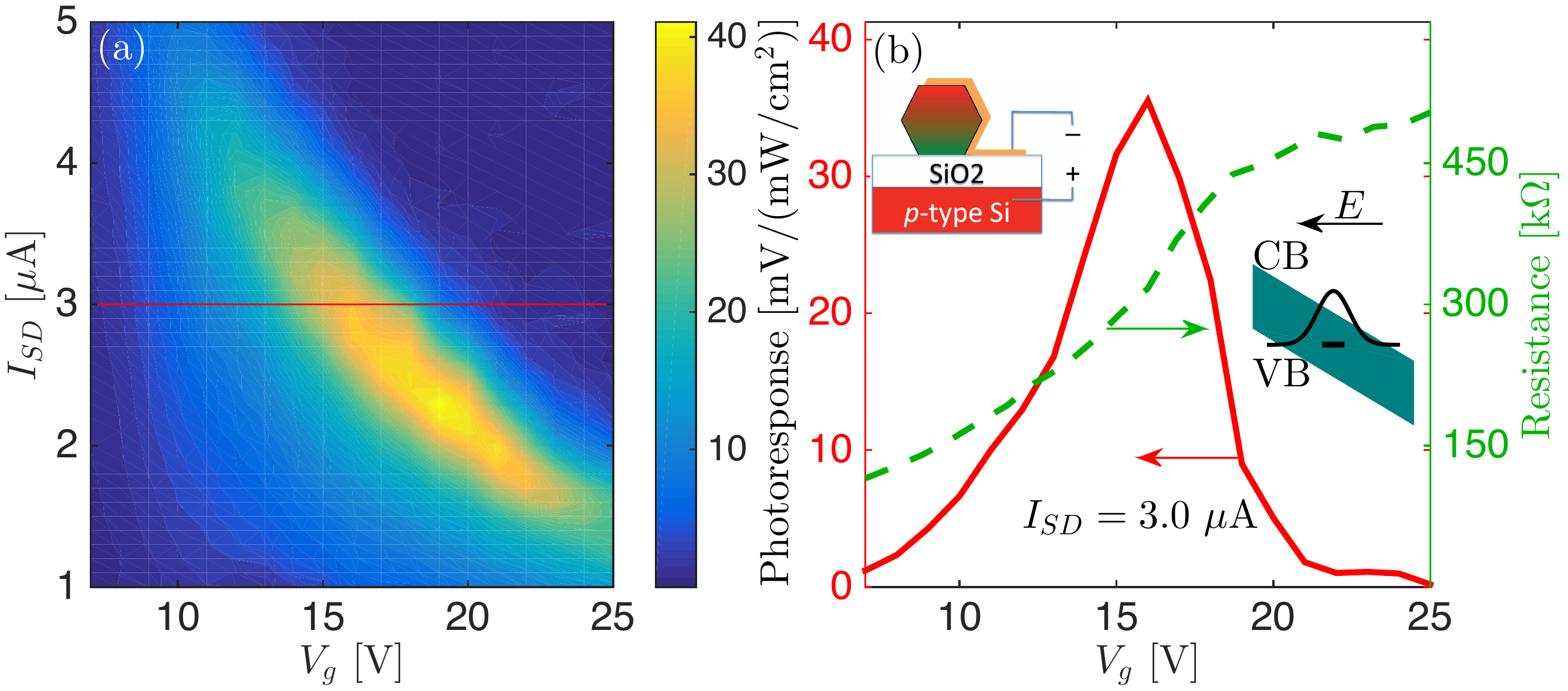}
\caption{Bias dependence of the photoconductivity of a 240~nm-thick TeNW device at 77K. 
(a) Contour of the `TE' photoresponse at 2.8 $\mu$m as a function of the back-gate bias $V_g$ and current bias $I_{SD}$. Panel (b) shows the data in (a) at constant bias current of 3~$\mu$A (solid red curve), together with the DC resistance (dashed green curve) as a function of $V_g$. The inset on the left shows the separation of the charge distributions of the electrons (in green) and holes (in red) under positive back gate bias. The inset on the right shows band tilting under strong electric field, which enhances the tunneling of free carriers into the recombination center.
\label{fig:bias_dep} }
\end{figure}

The application of back-gate bias on the TeNW devices provides an effective approach to control the photoresponse behavior via electric field. Similar to its temperature dependence, we find that the relative shape of the spectra is unaffected by the gate bias, whereas the amplitude of the photoresponse is strongly affected. 
Specifically, a positive gate bias voltage could boost the photoresponse signal by orders of magnitude, but beyond a certain threshold, larger bias suppresses the photoresponse, in contrast with the relatively weak tunability of the DC resistance under gate bias.   

As an example of this observation, Fig.~\ref{fig:bias_dep}(a) shows a contour of the photoresponse from a 240~nm thick TeNW sample under illumination by 2.8 $\mu$m TE polarized light at 77K, as a function of the positive back gate voltage and the DC current, showing strongly nonlinear features. As seen in Fig.~\ref{fig:bias_dep}(b) under $I_\text{SD} = 3$~$\mu$A, the photoresponse is very weak at low gate bias but increases superlinearly, reaching maximum around $V_g=17$~V, and then drops drastically with further increased gate bias. As a comparison, the DC resistance increases monotonically only by a factor of 4 within this range of gate bias. 
The bias-dependence of the photoresponse in the TeNW devices is governed by the interplay of multiple mechanisms, under which the recombination of excess carriers could be influenced by the applied electric field in opposite ways.

First of all, a positive electric field reduces the majority carrier density $p_0$ and enhances both the radiative and Auger lifetimes discussed in the previous section. Secondly, electric field from the back gate alters the local densities of the electrons and holes, inducing accumulation in opposite directions [see Fig.~\ref{fig:bias_dep}(b), inset on the left hand side]. As a macroscopic result, the spatial distributions of the two types of carriers overlap less, so that their mutual recombination is reduced. Both these two factors contribute to the enhancement of the photoresponse. However, the second mechanism would eventually saturate due to the finite thickness of the nanowire, since the separated two types of charges accumulate at the opposite sides and build in an internal electric field counteracting with the external one.

The suppression of the photoresponse at large enough bias could be understood via an opposing mechanism. Under external electric field, the conduction and valence bands are spatially tilted. As a result, the trap states as recombination centers within the gap are no longer strictly bound but delocalized in $k$-space due to bound-to-band tunneling transitions [see Fig.~\ref{fig:bias_dep}(b), inset on the right hand side], so that excess carriers are easier to be captured and annihilated. The enhancement to the trap-assist recombination at large electric field is exponential, especially at low temperature.\cite{Schenk_SSE1992}

\section{Mie theory Simulation}

The thickness of our TeNW samples is in the range of several hundred nanometers. Given the Te refractive indices $n_o = 4.9$ and $n_e=6.5$, mid-infrared light with free-space wavelength around 3~$\mu$m has a wavelength inside the nanowire comparable with the thickness. Such matching in dimension promotes resonance of the light inside the nanowire forming a cavity, and the ‘resonant-mode’ features could be observed by photoconductivity spectra as reported in nanowire devices of other types of semiconductor, such as Ge,\cite{Cao_NM2009} Si,\cite{Bronstrup_ACSnano2010} and InP,\cite{Grzela_thesis2013} in the visible and near-infrared ranges.

\begin{figure}[t]
\includegraphics[width=3.3in]{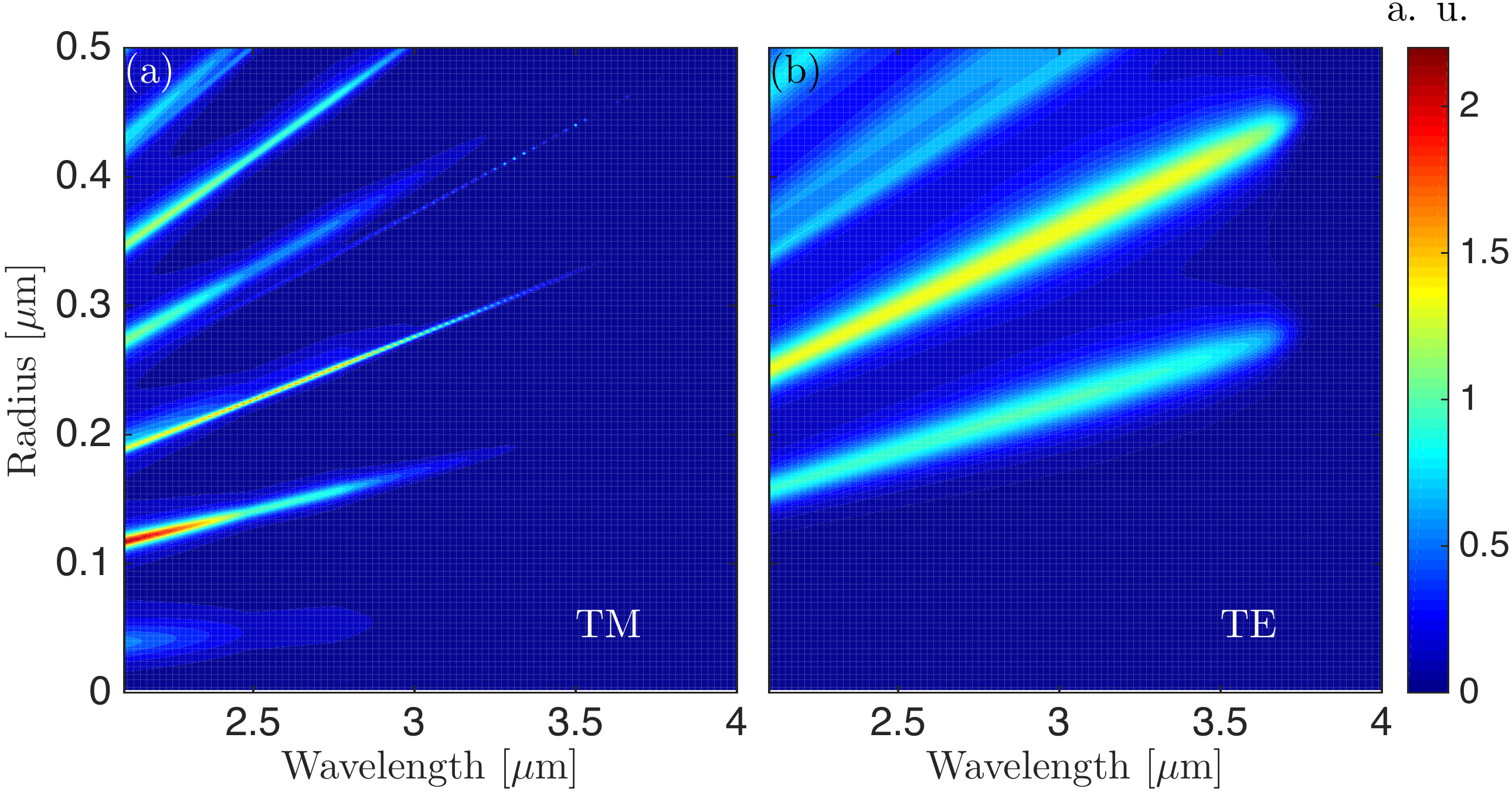}
\caption{Mie theory calculations of absorption spectra of cylindrical Te as functions of the wavelength and the radius, under the illumination of light with linear polarizations (a) along and (b) transverse to the nanowire axis.
\label{fig:Mie} }
\end{figure}

According to electromagnetic theory, the resonant spectra could be simulated by the formalism developed by Mie about light absorption and scattering by cylindrical shape nanostructure with radius $r$. 
The absorption efficiency for normal incident light with free-space wavevector $k_0$ is\cite{Mie_book}
\begin{align}
Q = \frac{2}{k_0r}\!\left[\text{Re}\!\left(\!a_0+2\sum_{n=1}^\infty a_n\!\right)-\left(\!|a_0|^2+2\sum_{n=1}^\infty |a_n|^2\!\right)\!\right],
\end{align}
where the unitless coefficients $a_n$s are given differently for TE and TM polarizations as
\begin{align}
a_n = \left\{\begin{array}{cc} \frac{J_n'(\bar{n}_ok_0r)J_n(k_0r)-\bar{n}_oJ_n(\bar{n}_ok_0r)J_n'(k_0r)}{J_n'(\bar{n}_ok_0r)H_n(k_0r)-\bar{n}_oJ_n(\bar{n}_ok_0r)H_n'(k_0r)}, & \text{for TE} \\ 
\frac{\bar{n}_eJ_n'(\bar{n}_ek_0r)J_n(k_0r)-J_n(\bar{n}_ek_0r)J_n'(k_0r)}{\bar{n}_eJ_n'(\bar{n}_ek_0r)H_n(k_0r)-J_n(\bar{n}_ek_0r)H_n'(k_0r)}, & \text{for TM} \\
\end{array}\right.
\end{align}
with $J_n(x)$ and $H_n(x)$  the cylindrical Bessel function and the first kind Hankel function, respectively. The complex refractive indices $\bar{n}_o = n_o+i\kappa_o$ (ordinary, for TE) and $\bar{n}_e = n_e+i\kappa_e$ (extraordinary, for TM) are wavelength-dependent, and the imaginary parts $\kappa_{o,e}$ are responsible for the cross-bandgap absorption, hence the photo-excitation of charge carriers.
In our case of TeNW, the input data of refractive indices of Te are taken from Ref.~[\onlinecite{PALIK_HB1997}]. The calculated absorption spectra of the TM and TE polarized light are shown in the contours in Fig.~\ref{fig:Mie}, as functions of the wavelength of the incident light and the radius of the nanowire. It is reasonable that the characteristic wavelength of each resonant mode scales linearly with the size of the nanowire. The band gap feature is obvious around the 3.5~$\mu$m (TM) and 3.7~$\mu$m (TE) cutoffs, due to the vanishing of $\kappa_{o,e}$ at longer wavelength. Also notice the much wider linewidths of TE modes in general, compared with those of TM, as evidence of the fact that $\kappa_o$ is roughly two orders of magnitude larger than $\kappa_e$, within the wavelength range of our focus,\cite{PALIK_HB1997} reflecting their different origins of cross-bandgap optical transitions as discussed in the introduction.

The observed photoresponse spectra of our TeNW devices, however, are not fully consistent with the theoretical expectation. Except sharing the ‘bandgap’ feature governed by the same kind of material, the shapes of the photo-response spectra, although varying from sample to sample with different thickness, do not show clear resonance features as in Fig.~\ref{fig:Mie}. It is confirmed that the measured spectra are robust against the length of the device: along the same nanowire, contact pairs with different spacings record similar shape of the photoresponse spectra, ruling out possible antenna effect.

There are several possible factors contributing to such a discrepancy. First, the surface of the nanowire, once exposed in ambient, is inevitably coated with an oxide layer to certain thickness, so that the $Q$ factor of the nanowire as a cavity is reduced, suppressing the resonant features. 
Moreover, scanning electron microscopy (SEM) shows that the cross sections of the nanowires are `squeezed' hexagons with varied sides. Some samples, especially the thick ones, are actually nanotubes with hollows at the center. Such complicated reality of the geometry of the samples, together with the external environment of Si/SiO$_2$ substrate (rather than in free space), may also contribute to the deviation from the ideal and simple consideration in the Mie formalism.


\section{summary}
We have presented a comprehensive study on polarization-resolved photoconductivity spectra of TeNW devices in the mid-infrared range. The temperature dependence of the amplitude of the photoresponse demonstrates domination of the trap-assisted and Auger recombination processes beyond 100K. A moderate value of positive gate bias could enhance the photoresponce by more than an order of magnitudes.

Our simulation based on the Mie theory suggests that the performance of the photoresponse on TeNW devices could be further improved by refining the quality of the samples. Surface coating on freshly synthesized samples, such as by atomic layer deposition of Al$_2$O$_3$, could be used to prevent surface oxidation and promote the resonant modes. In addition to gate tuning, applying proper n-type dopants to compensate the p-type carriers may be an alternative in  suppressing the residual conductivity and prolong the lifetime of the photo-excited carriers. Effective application of local doping could form p-n junction array along the nanowires for high resolution photovoltaic mid-infrared detection.

Taking the advantage of the nanowire geometry and the strong spin-orbit coupling in Te, the photo-excited carriers could be manipulated to study interesting spin-related physics. 
For example, two-beam orthogonal interference  produces optical grating of circularly polarization along the direction of the nanowire axis, which generates spin polarization of the hole carriers moving in one of the two directions along the nanowire, due to the special spin-momentum correlation of the valence band states.\cite{Asnin_SSC1978} 
Engineered such photo-generated spin polarization could be used to probe the spin texture of the theoretically predicted Weyl points.\cite{Hirayama_PRL2015}

\section*{Acknowledgements}
We thank Professor Ian Appelbaum for providing many helpful suggestions during this study. We acknowledge the support of the Maryland NanoCenter and its FabLab and AIMLab, and the UMD Surface Analysis Center. 
This study is supported by the Office of Naval Research under Contract No. N000141712994.

\section*{Data availability}
The data that support the findings of this study are available from the corresponding author
upon reasonable request.
\bibliography{TeNW_PC}

\end{document}